\documentstyle[12pt]{article}
\def\be{\begin{eqnarray}}
\def\ee{\end{eqnarray}}

\input{psfig.sty}

\begin{document}

\title{\bf Legendre expansion of the $\nu \bar{\nu} \rightleftharpoons 
e^+ e^-$ kernel: \\ Influence of high order terms.}

\author
{\bf J. A. Pons $^1,2$ , J. A. Miralles$^1$ and
J. M$^{\underline{\mbox{a}}}$ Ib\'a\~{n}ez$^1$  \\
$^1$ Departament d'Astronomia i Astrof\'{\i}sica. \\ Universitat de Valencia
46100 Burjassot (Valencia) Spain  \\
$^2$ Astronomy program, Department of Physics and Astronomy, \\ SUNY at
Stony Brook, Stony Brook, New York, 11794}

\maketitle

\noindent
To be published in {\it Astron. Astrophys. Suppl. Ser.} {\bf 129} (1998)

\begin{abstract}
We calculate the Legendre expansion of the rate of the process $\nu + \bar{\nu}
\leftrightarrow e^+ + e^-$ up to $3^{rd}$ order extending previous
results of other authors which only consider the $0^{th}$ and $1^{st}$ 
order terms. 
Using different closure relations for the moment equations of
the radiative transfer equation we
discuss the physical implications 
of taking into account quadratic and cubic terms
on the energy deposition outside the neutrinosphere in a simplified
model. The main conclusion is that $2^{nd}$ order is necessary
in the semi-transparent region and gives good results if an appropriate 
closure relation is used.  

\end{abstract}

\section{Introduction}

The neutrino emission processes play an important role in
different astrophysical scenarios, in particular during the
stellar core collapse and the cooling of a protoneutron star.
Recently, several authors have studied some processes
such as neutrino-electron scattering (Smit et al. \cite{Smi96} 
and Cernohorsky, J. \cite{Cer94}), neutrino
coherent scattering off nuclei (Leinson \cite{Lei92}),
effects of nucleon spin fluctuations in the weak
interaction rates (Janka et al. \cite{Jan96}) or neutrino reactions
with strange matter (Reddy and Prakash \cite{Red97}) in order to
obtain the rates to be used in transport calculations.
We will focus on the
thermal pair emission-absorption process 
$e^+ e^- \leftrightarrow \nu \bar{\nu}$ (TP in the next). 
In the stellar core collapse
scenario, discrepancies in the efficiency of TP in the heating
of matter outside the neutrinosphere have been noticed
(see, e.g., Janka, 1990 and references therein) and in order to
obtain meaningful calculations of the energy deposition rate it is
necessary to carefully consider
the angular dependence of the distribution function.

In some calculations involving neutrino transport, only the
$0^{th}$ and $1^{st}$ terms of the Legendre expansion of the collision
kernel are included as in Bruenn (\cite{Bru85}) and Suzuki (\cite{Suz89}). 
This approach could suffice where the diffusion 
approximation remains valid. The justification of that approach
relies on the fact that the $2^{nd}$ and $3^{rd}$ order
terms appear multiplying $2^{nd}$ and $3^{rd}$ order terms of the
Legendre expansion of the
neutrino distribution function ($I$) and they 
vanish in the diffusion limit. 
However, if a general closure relation is used which is
different to $P=\frac{1}{3}E$, these higher order
terms of $I$ do not vanish, and their contribution becomes 
especially important in the semi-transparent region.
In this paper we will also analyze the influence of
the different closure relations that have been considered in recent
years.

This work is organized as follows:
In \S2 we present the Legendre expansion of the TP production
and absorption kernels and we give
explicit expressions for the $0^{th}$ to $3^{rd}$ order terms. 
In \S3 we study how
these new terms affect the sources in the two-moment closure
transport equations. In \S4 we discuss the effects of the new terms
and the influence of the closure relation.

\section{Legendre expansion of emission-absorption TP kernels}

Following Bruenn (\cite{Bru85}) the contribution of thermal pair production
and absorption of neutrino--antineutrino pairs to the right
hand side of the Boltzmann equation (TP collision term) is 
\be
B_{TP}[I,\bar{I}]=
\frac{1}{c{(hc)}^3} \int_0^{\infty} \omega'^2
d\omega'\int_{-1}^{+1} d\mu'\int_0^{2\pi}d \phi
\nonumber \\
\left\{ [1-I][1-\bar{I}] R_{TP}^p(\omega,\omega',\cos\theta) -
I \bar{I} R_{TP}^a(\omega,\omega',\cos\theta) \right\}
\ee

\noindent where $I=I(t,\vec{r},\mu,\omega)$ ($\bar{I}=\bar{I}(t,\vec{r},\mu',\omega')$)
is the neutrino (antineutrino)  invariant distribution
function, $\omega$ ($\omega'$) is the neutrino (antineutrino)
energy in the frame comoving with the matter, $\mu$ ($\mu'$)
the cosine of the angle between the neutrino 
(antineutrino) momentum and the polar axis, $\phi$ is the azimuthal angle
and $\theta$ is the
angle between neutrino and antineutrino directions. 
In what follows, the explicit dependence on $t$ and $\vec{r}$ of 
the distribution functions will be omitted
and we will assume axial symmetry with respect to the polar axis.
The superscripts $a$ and $p$ refer to absorption and production,
respectively.

Assuming electrons and positrons in equilibrium
at a temperature $T$, the following
relation between the absorption and production kernels is satisfied
\be
R_{TP}^a(\omega,\omega',\cos\theta)=e^{\frac{\omega+\omega'}{T}}
R_{TP}^p(\omega,\omega',\cos\theta)
\ee
where the temperature is measured in units of energy.
Expanding $R_{TP}^p$ as follows
\be
R_{TP}^p(\omega,\omega',\cos\theta)=\sum_l \frac{2l+1}{2}
\Phi_l(\omega,\omega') P_{l}(\cos\theta)
\ee
where $P_l(\cos\theta)$ are the Legendre polynomials and 
$\Phi_l(\omega,\omega')$
are the Legendre moments of the production kernel,
the TP collision term can be written as
\be 
B_{TP}[I,\bar{I}]=\frac{2\pi}{c{(hc)}^3}\int_0^{\infty} \omega'^2
d\omega' 
\nonumber \\
\left\{ (1-I)\Phi_0 - \sum_l \frac{2l+1}{2} 
\Phi_l  P_l(\mu)
\int_{-1}^{+1} d\mu' P_l(\mu') \bar{I} \right. + 
\nonumber \\
\left. \left( 1-e^{\frac{\omega+\omega'}{T}} \right) \sum_l \frac{2l+1}{2}
\Phi_l P_l(\mu) I \int_{-1}^{+1} d\mu' 
P_l(\mu')  \bar{I} \right \}
\ee

In order to obtain the expressions for $\Phi_l$, 
we need to evaluate the following integral over electron energy $E$
\be
\Phi_l=\frac{G^2}{\pi} \int_0^{\omega+\omega'} dE F_{e}(E,\eta)
F_{e}(\omega+\omega'-E,-\eta)  \times
\nonumber\\
\left[ \alpha_1^2 J_l(\omega,\omega',E) +
\alpha_2^2 J_l(\omega',\omega,E) \right]
\ee
where $\alpha_1$ and $\alpha_2$ are summarized in table 1 for the
different neutrino types and $F_{e}(E,\eta)$ is the Fermi-Dirac
distribution function

\be 
{\displaystyle F_{e}(E,\eta)=\frac{1}{e^{\frac{E}{T}-\eta}+1} }
\ee
$\eta$ being the electron degeneracy parameter,
and 
\be 
J_l(\omega,\omega',E)=\frac{1}{\omega \omega'} \Theta(\omega+\omega'-E) \times
\nonumber \\
\int_{-1}^{+1}  d\mu P_l(\mu) \left[ A(\mu)+B(\mu)E
+C(\mu)E^2 \right] \Theta(\mu-\mu_0)
\ee

\be 
\mu_0 = 1-\frac{2E(\omega+\omega'-E)}{\omega \omega'} 
\ee
In the previous expression $\Theta$ stands for the step function.

Although $A(\mu)$, $B(\mu)$ and $C(\mu)$  
are complicated
functions of $\mu$, $\omega$ and $\omega'$, the integrals $J_l$
can be done analytically, and the results for $l=0,1$ were first
obtained by Bruenn (\cite{Bru85}). We have calculated the
integrals for $l=2,3$ and 
their dependence on $E$ is simply a polynomial law of degree $2l+5$
with coefficients being functions of $\omega$ and $\omega'$.

Then, taking into account the following relation
\be
F_{e}(E,\eta)F_{e}(\omega+\omega'-E,-\eta) = 
\nonumber\\
\frac{1}{1-e^{\frac{\omega+\omega'}{T}}} \left[ 
F_{e}(E,\eta) - F_{e} \left( E,\eta+\frac{\omega+\omega'}{T} \right) \right]
\ee
the $\Phi_l$ functions can be expressed in a simple way
in terms of the dimensionless variables $y=\omega/T$ and
$z=\omega'/T$.

\be 
\Phi_l(y,z)=\frac{G^2}{\pi} \frac{T^2}{1-e^{(y+z)}} 
\left[ \alpha_1 \Psi_l(y,z) + \alpha_2 \Psi_l(z,y) \right] 
\ee

\be
\Psi_l(y,z) = 
\sum_{n=0}^{2} \left( 
c_{ln} G_n(y,y+z)+
d_{ln} G_n(z,y+z) \right) 
\nonumber \\
+ \sum_{n=3}^{2l+5} a_{ln} \left(
G_n(0,\mbox{min}(y,z)) - G_n(\mbox{max}(y,z),y+z)\right) 
\ee
where 
\be
G_n(a,b)=\int_a^b dx \frac{x^n}{e^{x-\eta}+1} - \int_a^b dx
\frac{x^n}{e^{x-(\eta+y+z)}+1}
\ee

The explicit expressions for $a_{ln}$, $c_{ln}$ and $d_{ln}$ coefficients are in
Appendix A and the method to evaluate the $G_n(a,b)$ integrals
is detailed in Appendix B.

\section{Energy and momentum transfer in two moment neutrino transport}

Two moment neutrino transport (Cernohorsky and van Weert \cite{Cer92})
consists in solving the spectral energy and momentum balance equations
as a coupled set.
Let us define the $l^{th}$ moment of the distribution function 
\be
I_l(\omega) = \frac{1}{2} \int_{-1}^{+1} I \mu^{l} d\mu
\ee
and the following {\it Eddington factors}

\be
f(\omega) = \frac{I_1(\omega)}{I_0(\omega)}
\ee
\be
p(\omega) = \frac{I_2(\omega)}{I_0(\omega)}
\ee
\be
q(\omega) = \frac{I_3(\omega)}{I_0(\omega)}
\ee
For simplicity we will omit the energy dependence.
Quantities with bar ($\bar{I}$) will stand for antineutrinos
and analogous definitions to (13-16) will be used.

In order to close the set formed by the two equations 
(energy and momentum) we need two
closure relations $p=p(f,I_0)$ and $q=q(f,I_0)$. In the next section
we give some closure relations widely used in the literature. 

The source terms 
in the energy and momentum balance equations are
obtained by angular integration of the 
collision term on the right hand side of the Boltzmann equation 

\be
{\left( \frac{\partial I_0}{\partial t} \right)}_{TP}
= \frac{1}{2} \int_{-1}^{1} d\mu B_{TP}
= \frac{2\pi}{c{(hc)}^3}\int_0^{\infty} 
\omega'^2 \, d\omega' \, S_0(\omega,\omega')
\ee
                                                       
\be
{\left( \frac{\partial I_1}{\partial t} \right)}_{TP}
= \frac{1}{2} \int_{-1}^{1} d\mu \mu B_{TP}
= \frac{2\pi}{c{(hc)}^3}\int_0^{\infty} 
\omega'^2 \, d\omega' \, S_1(\omega,\omega')
\ee                                                    

These source terms are the contribution of the TP
process to the energy and momentum exchange between neutrinos and matter.
After using the Legendre expansion (4) the expression for
$S_0$ and $S_1$ can be obtained

\be
{S_0}= \left[ 1-I_0-\bar{I}_0 +
\left( 1-e^{\frac{\omega+\omega'}{T}} \right) \Gamma_0(\omega,\omega') 
\right] \Phi_0
\ee
\be
{S_1}=
- I_1 \Phi_0  - \bar{I}_1 \Phi_1 
+ \left( 1-e^{\frac{\omega+\omega'}{T}} \right) \Gamma_1(\omega,\omega')
\ee

\be
\Gamma_0 = \sum_{l} \frac{2l+1}{4} \frac{\Phi_l}{\Phi_0}
\int_{-1}^{1} d\mu P_l(\mu) \, I
\int_{-1}^{1} d\mu' P_l(\mu') \, \bar{I}
\ee
                                                       
\be
\Gamma_1 = \sum_{l} \frac{2l+1}{4} {\Phi_l}
\int_{-1}^{1} d\mu P_l(\mu) \, \mu \, I
\int_{-1}^{1} d\mu' P_l(\mu') \, \bar{I}
\ee

\section{Discussion}

If neutrino transport calculations are done in the 
diffusion approximation, where the distribution function is 
truncated at first order
\be
I(\mu,\omega) = I_0(\omega) + 3 \mu I_1(\omega)
\ee
the expansion in (21) is also truncated at the same order
and, hence,  there is no need to calculate high order terms of the kernels.
However, in the
semitransparent region where the diffusion approximation breaks down
and the flux factor is big,
one needs to use a closure relation which is different from
$P=\frac{1}{3}E$ and
the terms in $\Phi_l$ for $l\ge2$ must be taken
into account because, as we will demonstrate,  
keeping only the first order term 
would give a wrong value of the energy exchange between neutrinos
and matter.
To see this fact more clearly, let us study the influence on 
the energy source of $2^{nd}$ and $3^{rd}$ order terms truncating
$\Gamma_0(\omega,\omega')$ at $l=3$.

\be
\Gamma_0(\omega,\omega') = 
1 +  \left\{ f \bar{f} \right\} 3 \frac{\Phi_1}{\Phi_0} 
+ \left\{\frac{(3p -1)}{2}\frac{(3 \bar{p} -1)}{2} \right\} 
5 \frac{\Phi_2}{\Phi_0} 
+ \left\{\frac{(5q-3f)}{2}\frac{(5\bar{q} -3\bar{f})}{2} \right\} 
7 \frac{\Phi_3}{\Phi_0}
\ee

In this expression we have written inside brackets $\left\{ \right\}$
the part of the correction due to the distribution function. The value of
these factors is restricted to the interval $[0,1]$. Therefore, the
maximum value of each new term is $(2l+1)\frac{\Phi_l}{\Phi_0}$, which
we have plotted in figures 1a,1b and 1c for $l=1,2,3$, respectively.
As we can see from the plots, if terms in brackets are not small, 
the $2^{nd}$ contribution should be included for all energies and
the $3^{rd}$ order term has a significant contribution for low
energies. Only when the terms in brackets are much smaller than 1 the 
expansion can be truncated at first order.

The closure relation becomes, therefore, a fundamental point in 
the calculation of the $e^{+}e^{-}
\leftrightarrow \nu \bar{\nu}$ emission-absorption rate.
There are several closure relations used by different authors
and the question 'which is the best one?' has no answer yet.
For the sake of comparison, in this work we include four different
closure relations: MB (Minerbo \cite{Min78}) , 
LP (Levermore and Pomraining \cite{Lev81}),
CB (Cernohorsky and Bludman \cite{CB94}) and MH (Mihalas \cite{Mih84}). 
Cernohorsky closure depends on both, the $0^{th}$
and $1^{st}$ moments of the distribution function $p=p(I_0,f)$ and
$q=q(I_0,f)$ and the rest of them
are uniparametric closures, this is, $p=p(f)$ and 
$q=q(f)$. In the small occupation limit ($I_0 \ll 1$) CB closure
is equivalent to Minerbo's one and in the maximal forward angular
packing limit is equivalent to the {\it vacuum approximation} closure (see below).

The form of these closures is the following:
\be
MB \,\,\, \left\{ {\matrix{
f(a)=\mbox{coth}(a)-1/a \cr
p(a)=1-2f(a)/a \cr
q(a) = f(a) - (3p(a)-1)/a \cr}} \right.
\ee
\be 
LP \,\,\, \left\{ {\matrix{
f(a)=\mbox{coth}(a)-1/a \cr
p(a)=\mbox{coth}(a) f(a) \cr
q(a)=\mbox{coth}(a) p(a) - 1/3a \cr}} \right.
\ee
\be 
MH \,\,\, \left\{ {\matrix{
p(f)=(1+2f^2)/3 \cr
q(f)=(3f+2f^3)/5 \cr}} \right.
\ee
\be 
CB \,\,\, \left\{ {\matrix{
p(I_0,f)={\frac{1}{3}+\frac{2}{3}(1-I_0)(1-2I_0)
\chi\left( \frac{f}{1-I_0} \right)} \cr
q(I_0,f)= \mbox{non-analytic} \cr}} \right.
\ee
where $\chi(x)=1-3x/\beta(x)$ with $\beta(x)$ being the inverse of the Langevin
function $x=\coth \beta - 1/\beta$. The following expression fits this function well
(Cernohorsky and Bludman \cite{CB94}) 
$$ \chi(x) = \frac{x^2(3-x+3x^2)}{5} $$

In figure 2 we show $p(f)$ and $q(f)$ for the different
closure relations, taking an occupation level for the CB case of $I_0=0.1$.
In figure 3, the combinations of the different moments
that appear inside brackets in (24) are plotted as a function of $f$. In both 
figures the solid line is for CB closure, dotted line for MH,
dashed line for MB and dashed-dotted line for LP. We also plot
with crosses the closure obtained in the vacuum approximation.
As can be seen from the figures, there are important differences between 
different closures for $f\ge 0.3$. This can lead to relevant differences
in the energy exchange between neutrinos and matter for large values
of the flux factor $f$.

To illustrate this feature, let us study a simple model consisting 
of a sphere of radius $R$ radiating neutrinos and antineutrinos 
isotropically into vacuum, the so called {\it vacuum approximation}
(Cooperstein et al. \cite{Coo86}). The closure consistent 
with this model (VA) is
\be
VA \,\,\, \left\{ {\matrix{
f(a) = \displaystyle{\frac{1+a}{2}} \cr
p(a) = \displaystyle{\frac{1+a+a^2}{3}} \cr
q(a) = \displaystyle{\frac{1+a+a^2+a^3}{4}} \cr}} \right.
\ee

We assume that the neutrino (and antineutrino) spectrum 
at the surface of the sphere is
Fermi-Dirac with zero chemical potential and $T_{\nu}=1$ MeV. 
Of course in a real case the 
neutrinospheres of neutrinos and antineutrinos are located in
different places for different energies, but this example is
illustrative of the general behaviour of the heating rate. We 
calculate the net (heating minus cooling)
heating rate of the matter per unit volume for given  
distance to the center of the sphere ($d$)
and matter temperatures ($T$). 
For $T>T_{\nu}$ cooling dominates over heating and we find
that the effect of
including new orders is negligible. In contrast, for low $T$ the 
dominant term in eq. (19) is the term proportional to $\Gamma_0$ and there
are remarkable differences in the heating rates obtained including
the new terms. These effects are also closure dependent and we
compare the closure relations discussed previously in order to estimate
the differences between them. 

In figure 4 we present the result of
our calculations. We fix the matter temperature at $T=0.5 MeV$ and
we study the influence of higher order terms and closures for
different distances from the center of the star (different flux
factors). For the sake of comparison, we have performed Monte Carlo
integration of the complete expression for the reaction rate,
shown in the figure as the solid line. We overplot the total energy
deposition after including each new order, using different
symbols for the different closures. As can be seen, at first order
there is an underestimation of the deposition rate that is
worse for high flux factors, even changing the sign (this means
emission instead absorption of energy) for $f>0.8$. The inclusion of
the second order term gives a big improvement if one uses the closure  
consistent with the form of the distribution function in the model.
The third order term is a small correction that can be omitted
in all cases unless high accuracy is required. Using different
closures instead of VA gives worse results at small
$x$, but solves the problem in the sign for high values of $x$.
We also observe remarkable differences between the results obtained using
different closures, even though, we cannot deduce which one would have
better behaviour in a realistic case.

To summarize, we restate our main conclusions.
The first conclusion is that, in the semi-transparent region and
for matter temperature lower than neutrino temperature,
it is necessary to consider the expansion up to $2^{nd}$ order.
This procedure gives good results when combined with an appropriate 
closure relation. The $3^{rd}$ order term does not lead 
to a substantial improvement
in the solution
since, as we have shown, it is only a small correction.
For $T>T_{\nu}$
cooling dominates over heating and we find that the effect of
including new orders is negligible.

The second conclusion is that, even though convergence is reached
with $2^{nd}$ order corrections, the result is very sensitive to the
closure relation chosen. 
Best results are obtained 
when using a closure relation consistent with
the particular distribution function used in the model
and therefore, detailed study of the closure in each particular
problem is needed to obtain good estimates of interaction rates
in a multigroup flux-limited diffusion problem.
These results can be applied in all problems concerning neutrino
transport such as stellar core collapse or cooling of newly born
neutron stars.

\subsection*{Acknowledgements}
This work has been supported by the 
Spanish DGCYT grant
PB94--0973 and partially by the US Dept. of Energy grant DE-AC02-87ER40317.
We thank A. P\'erez for carefully reading the manuscript and
J.A.P. thanks J. Lattimer, M. Prakash and S. Reddy for useful discussions.

\newpage

\appendix

\section*{Appendix A: Expressions of $a_{ln}$, $c_{ln}$ 
and $d_{ln}$ coefficients}

The coefficients for $l=0,1$ are
the same as in Bruenn (\cite{Bru85}) but adapted to our notation.

\bigskip

{\bf Coefficients for $l=0$}

$$ c_{00} = \frac{4y}{z^2}(\frac{2z^2}{3}+{yz}+\frac{2y^2}{5}) $$
$$ d_{00} = \frac{4z^3}{15y^2}$$

$$ c_{01} = -\frac{4y}{3z^2}(3y+4z)$$ 
$$ d_{01} = -\frac{4z^2}{3y^2} $$

$$ c_{02} = \frac{8y}{3z^2}$$
$$ d_{02} = \frac{8z}{3y^2}$$

$$ a_{03} =  \frac{8}{3y^2}$$
$$ a_{04} = - \frac{4}{3y^2z}$$
$$ a_{05} =  \frac{4}{15y^2z^2}$$

{\bf Coefficients for $l=1$}

$$ c_{10} =  - \frac{4y}{z^3}(\frac{2}{7}y^3 + \frac{4}{5} y^2z +
\frac{4}{5} yz^2+ \frac{1}{3}z^3 )$$
$$ d_{10} = - \frac{4z^3}{105y^3}(14y+9z) $$

$$ c_{11} = \frac{4y}{z^3}(\frac{4}{5} y^2+ \frac{7}{5}yz +
\frac{2}{3}z^2) $$
$$ d_{11} =  \frac{4z^2}{5y^3}( \frac{7}{3}y+2z)$$

$$ c_{12} = - \frac{4y}{z^3}(\frac{3}{5}y+\frac{1}{3}z)$$
$$ d_{12} = - \frac{4z}{y^3}(\frac{1}{3}y+\frac{3}{5}z)$$

$$ a_{13} = \frac{8}{3y^2}$$
$$ a_{14} =- \frac{4}{3y^3z}(4y+3z)$$
$$ a_{15} = \frac{4}{15y^3z^2}(13y+18z)$$
$$ a_{16} = - \frac{4}{5y^3z^3}(y+3z)$$
$$ a_{17} = \frac{16}{35y^3z^3}$$

{\bf Coefficients for $l=2$}

$$ c_{20} = \frac{4y}{z^4}(\frac{2}{7}y^4+\frac{6}{7}y^3z
+ \frac{32}{35}y^2z^2 +  \frac{2}{5}yz^3 + \frac{1}{15}z^4) $$
$$ d_{20} = \frac{4z^3}{105y^4}(16y^2 + 27yz +12z^2)$$

$$ c_{21} = -\frac{4y}{z^4}(\frac{6}{7}y^3+\frac{12}{7}y^2z +
yz^2 + \frac{2}{15}z^3)$$
$$ d_{21} =  - \frac{4z^2}{y^4}
(\frac{18}{35}z^2 + \frac{6}{7}yz + \frac{1}{3}y^2)$$

$$ c_{22} = \frac{8y}{5z^4}(\frac{1}{6}z^2+\frac{3}{2}yz+\frac{12}{7}y^2) $$
$$ d_{22} =  \frac{8z}{5y^4}(\frac{1}{6}y^2+\frac{3}{2}yz+\frac{12}{7}z^2) $$

$$ a_{23} = \frac{8}{3y^2} $$
$$ a_{24} = -\frac{4}{3y^3z}(10y+9z) $$
$$ a_{25} = \frac{4}{15y^4z^2}(73y^2+126yz+36z^2) $$
$$ a_{26} = -\frac{12}{y^4z^3}(y^2+3yz+\frac{8}{5}z^2) $$
$$ a_{27} = \frac{48}{35y^4z^4}(2y^2+13yz+12z^2) $$
$$ a_{28} = -\frac{24}{7y^4z^4}(y+2z) $$
$$ a_{29} = \frac{8}{7y^4z^4} $$

{\bf Coefficients for $l=3$}

$$ c_{30} = -\frac{4y^2}{z^5}(\frac{10}{33}y^4+\frac{20}{21}y^3z
+ \frac{68}{63}y^2z^2 +  \frac{18}{35}yz^3 + \frac{3}{35}z^4) $$
$$ d_{30} = -\frac{4z^3}{105y^5}(9y^3 + 34y^2z + 40yz^2 +
\frac{500}{33}z^3)$$

$$ c_{31} = \frac{4y^2}{z^5}(\frac{20}{21}y^3+\frac{130}{63}y^2z 
+ \frac{48}{35}yz^2 + \frac{9}{35}z^3)$$
$$ d_{31} = \frac{4z^2}{35y^5}
(3 y^3 + 24y^2z + \frac{130}{3}yz^2 + \frac{200}{9}z^3)$$

$$ c_{32} = -\frac{4y^2}{z^5}
(\frac{50}{63}y^2+\frac{6}{7}yz+\frac{6}{35}z^2) $$
$$ d_{32} = -\frac{4z^2}{y^5}
(\frac{50}{63}z^2+\frac{6}{7}zy+\frac{6}{35}y^2) $$

$$ a_{33} = \frac{8}{3y^2} $$
$$ a_{34} = -\frac{4}{3y^3z}(19y+18z) $$
$$ a_{35} = \frac{4}{15y^4z^2}(253y^2+468yz+180z^2) $$
$$ a_{36} = -\frac{8}{15y^5z^3}(149y^3+447y^2z+330yz^2+50z^3) $$
$$ a_{37} = \frac{8}{21y^5z^4}(116y^3+\frac{2916}{5}y^2z+696yz^2+200z^3) $$
$$ a_{38} = -\frac{40}{21y^5z^5}(5y^3+54y^2z+108yz^2+50z^3) $$
$$ a_{39} = \frac{40}{21y^5z^5}(10y^2+43yz+\frac{100}{3}z^2) $$
$$ a_{3(10)} = -\frac{40}{9y^5z^5}(3y+5z) $$
$$ a_{3(11)} = \frac{320}{99y^5z^5}$$

\newpage

\section*{Appendix B:Evaluation of the $G_n$ integrals}
The integrals $G_n(a,b)$ appearing in the expression of the 
Legendre moments can easily be expressed as sums or differences
of the Fermi-like integral
$$F_n(\eta,x_1)=\int_0^{x_1} dx \frac{x^n}{e^{x-\eta}+1}$$
and therefore, the problem is reduced 
to calculate this kind of integrals.

In order to do this, we expand the denominator in the previous expression 
(Sack \cite{Sac90}), which must
be done in a different way depending on
the sign of $a=x-\eta$, this is :
$$ \frac{1}{e^a+1}= \left\{ {\matrix{
{\displaystyle \sum_{m=0}^{\infty}} {(-1)}^m e^{ma} & a<0 \cr
{\displaystyle \sum_{m=0}^{\infty}} {(-1)}^m e^{-(m+1)a}  & a>0 \cr}} \right. $$

\bigskip

\noindent then, we obtain an infinite sum of integrals that can be calculated
analytically using
$$ \int x^k e^{mx} = k! \sum_{l=0}^{k} \frac{{(-1)}^{k-l}}{l!}
\frac{e^{mx} x^l}{ m^{k+1-l}}$$
$$ \int x^k e^{-mx} = - k! \sum_{l=0}^{k} \frac{1}{l!}
\frac{e^{-mx} x^l}{m^{k+1-l}}$$

\bigskip

\noindent Let us define the following function
$$ T_l(\alpha) = \sum_{n=1}^{\infty} \frac{{(-1)}^{n+1} e^{-n\alpha}}{n^l} $$
which is well defined for $\alpha > 0$ and $l \geq 1$, we finally arrive
to a useful expression for the $F_k(\eta,x_1)$ integrals depending on
the value of $\eta$.
\bigskip

\noindent If $\eta<0$ 
\be F_k(\eta,x_1) = k! \left[ T_{k+1}(-\eta) - {\displaystyle \sum_{l=0}^k} 
\frac{T_{k+1-l}(x_1-\eta) x_1^l}{l!} \right] \nonumber \ee

\noindent If $0 \le \eta \le x_1$
\be
F_k(\eta,x_1) = \frac{\eta^{k+1}}{k+1} + 
k! \left[ 2 {\displaystyle \sum_{l=0}^{INT \left[\frac{k-1}{2} \right] }} 
\frac{T_{2l+2}(0) \eta^{k-1-2l}}{(k-1-2l)!} + \right.
\nonumber \\
\left. {(-1)}^{k} T_{k+1}(\eta) 
- {\displaystyle \sum_{l=0}^k} \frac{T_{k+1-l}(x_1-\eta) x_1^l}{l!} \right] 
\nonumber
\ee

\noindent If $x_1<\eta$
\be 
F_k(\eta,x_1) = \frac{x_1^{k+1}}{k+1} +  
k! \left[ {(-1)}^{k} T_{k+1}(\eta) - \right.
\nonumber \\  
\left. {\displaystyle \sum_{l=0}^{k}} {(-1)}^{k-l}
\frac{T_{k+1-l}(\eta-x_1) x_1^l}{l!} \right]  
\nonumber
\ee

\bigskip

The previous expressions are exact, and we only have to calculate 
a sum of a finite number of terms (up to $k$). The accuracy depends 
exclusively on the evaluation of the $ T_l(\alpha)$ functions. The
fact that these are uniparametric functions allows us to tabulate them
in a fine grid at the beginning of the calculation 
and to obtain enough accuracy without excessive CPU time cost.

\newpage

\begin{table*}
\begin{center}
\begin{tabular}{ccccc}
\hline \hline
$\, $ & $\nu_e \bar{\nu_e}$ & $\nu_{\mu,\tau} \bar{\nu}_{\mu,\tau}$ \\
\hline
$\alpha_1 $ & ${1+2\sin^2\theta_w}$ & $-1+2\sin^2\theta_w$  \\
$\alpha_2 $ & $2\sin^2\theta_w$ & $2\sin^2\theta_w$  \\
\hline \hline
\end{tabular}
\end{center}
\caption[]{Coefficients $\alpha_i$ for different neutrino species}
\end{table*}

\newpage

\begin{figure}
\psfig{figure=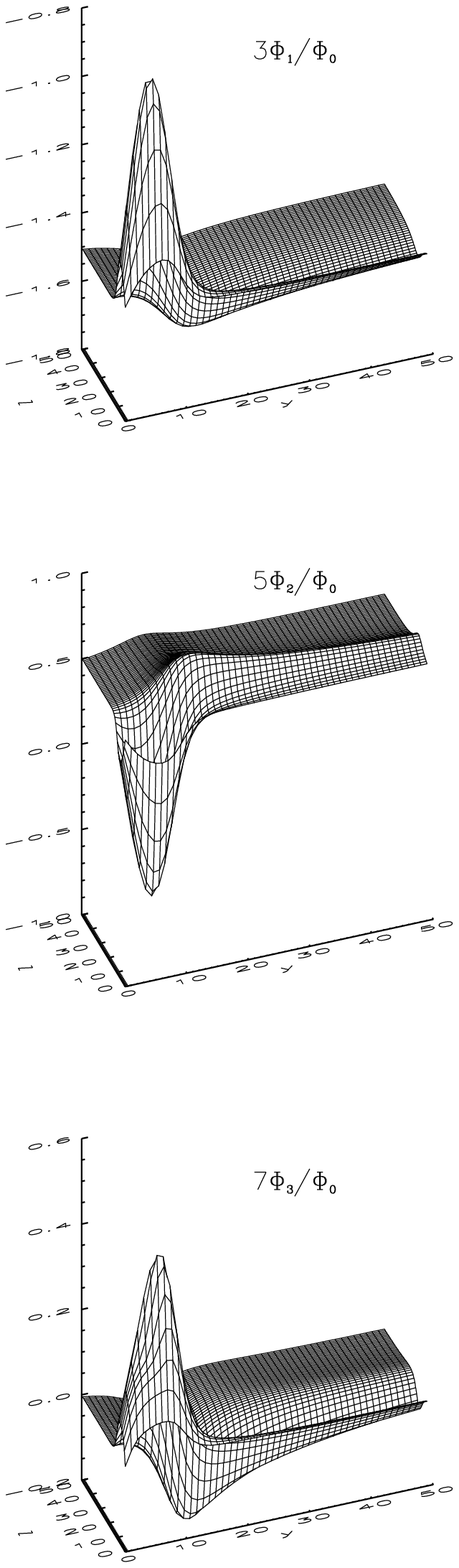}
\caption{Ratios $(2i+1)\Phi_i / \Phi_0$ for $i=1,2,3$ 
as functions of $y=\frac{\omega}{T}$ and $z=\frac{\omega'}{T}$ 
for an electron degeneracy parameter $\eta_e=10$ 
and $\mu$--$\tau$ neutrino type.}
\end{figure}

\newpage

\begin{figure*}
\psfig{figure=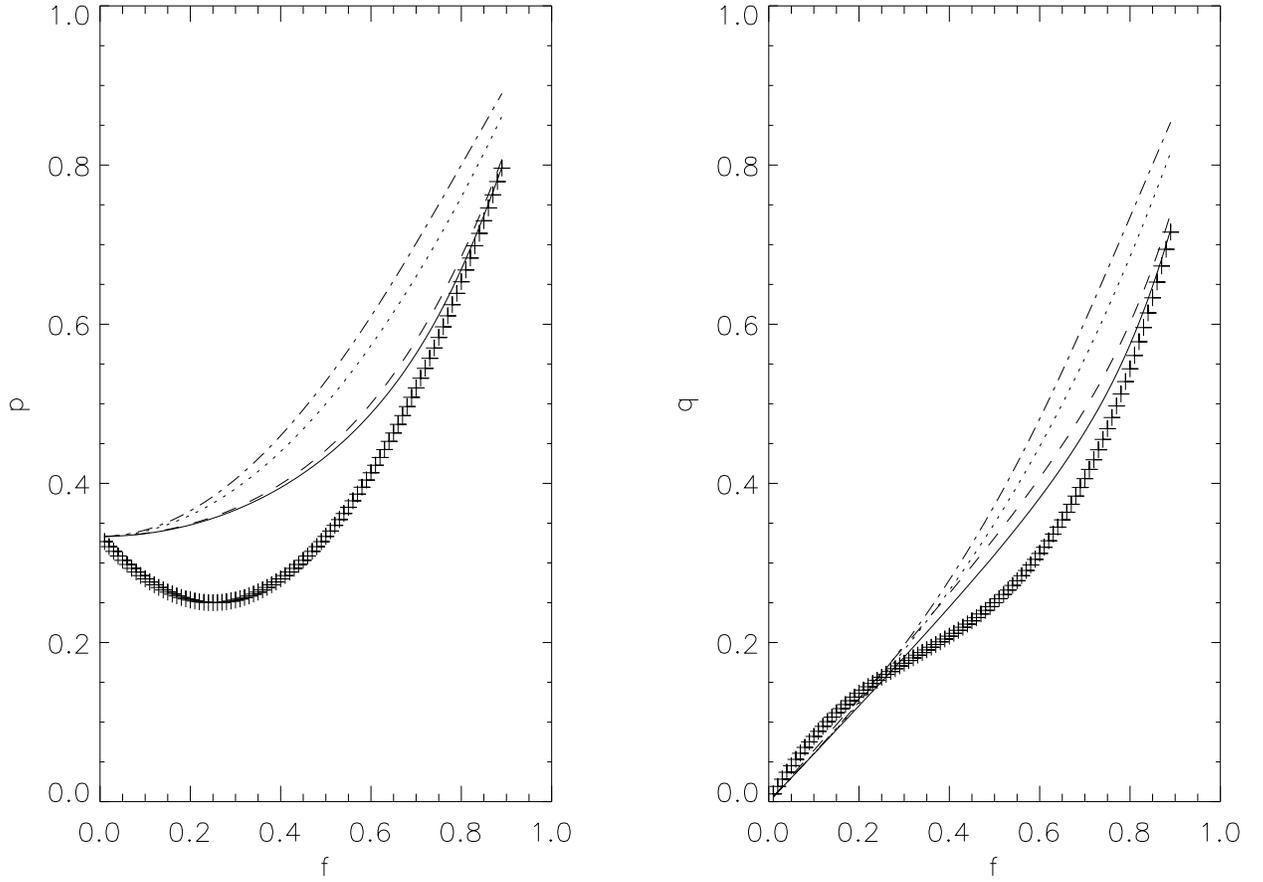}
\caption{Plots of  $p(f)$ and $q(f)$ for different closure
relations. The solid line is for CB closure, dotted line for MH,
dashed line for MB and dashed-dotted line for LP. We also plot
with crosses the closure obtained in the vacuum approximation.}
\end{figure*}

\newpage

\begin{figure*}
\psfig{figure=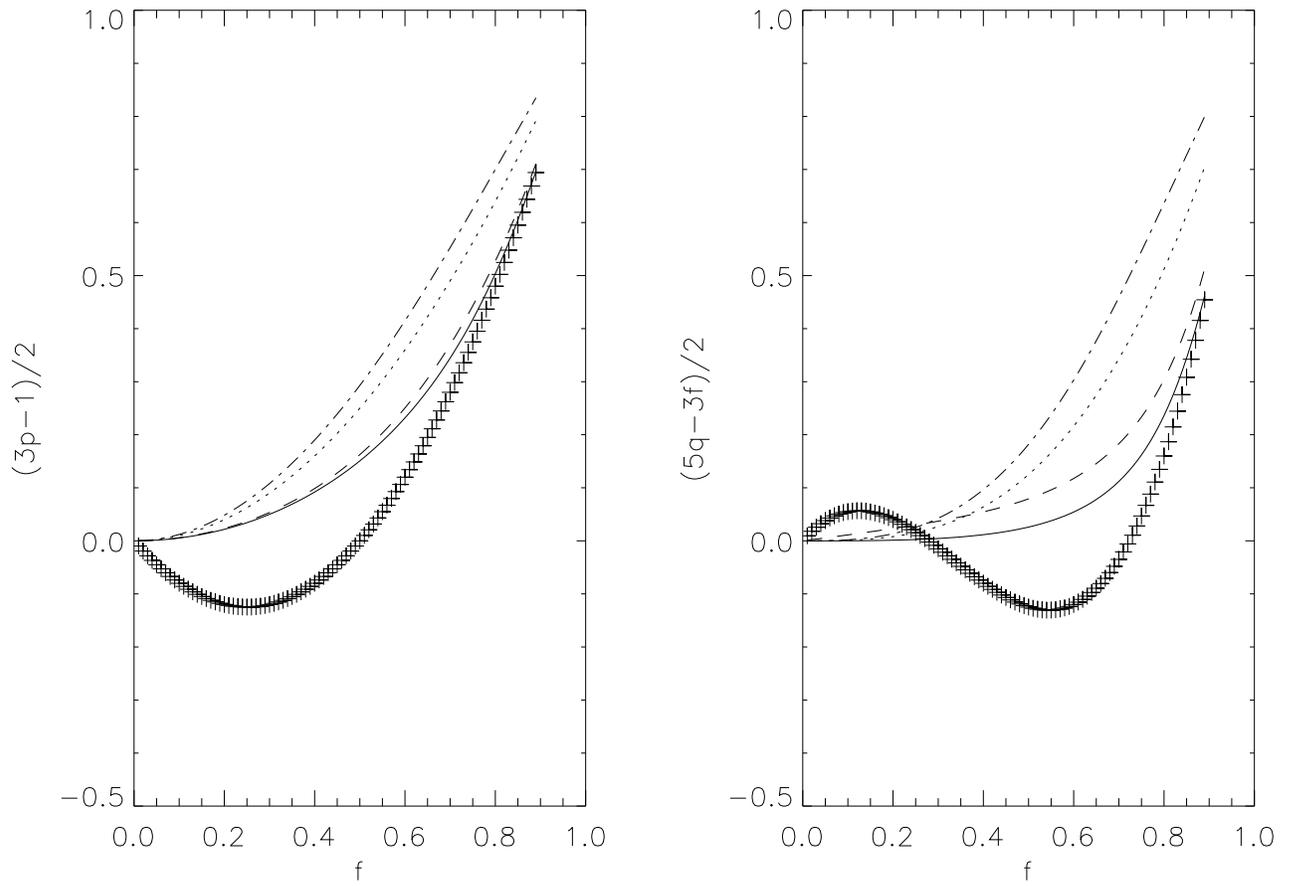}
\caption{Terms in brackets in the $2^{nd}$ and $3^{rd}$ order
contributions in the Legendre expansion  for different closures. The meaning
of the lines is the same as in Figure 2.}
\end{figure*}

\newpage

\begin{figure*}
\psfig{figure=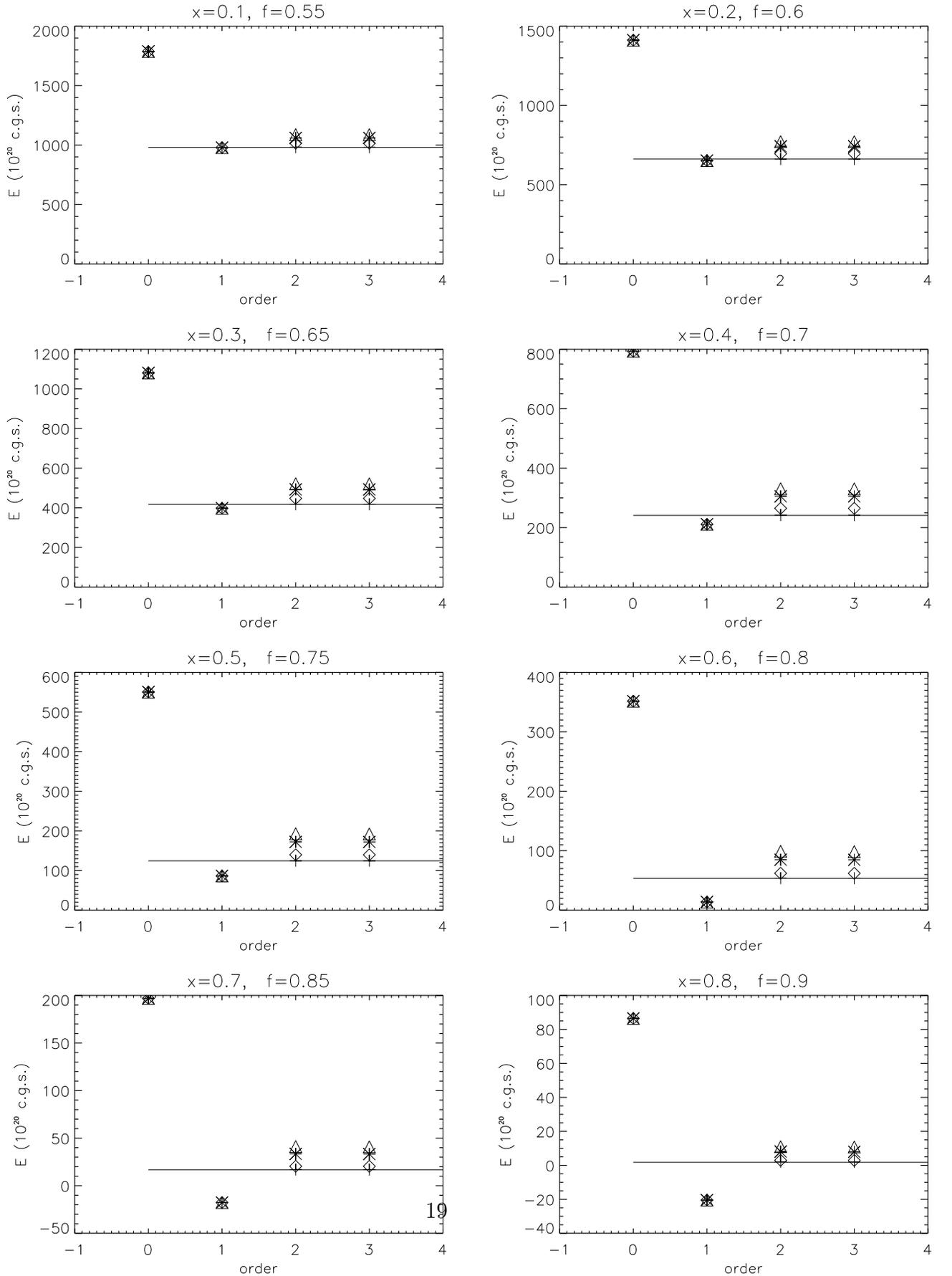}
\caption{Total energy deposited for different values of 
$x=\sqrt{1-{(R/d)}^2}$ in units of
$10^{20}$ \mbox{erg} ${\mbox{cm}}^{-3}$ ${\mbox{s}}^{-1}$. The solid line
is the exact solution after numerical integration and the different symbols
stand for the closures LP (triangle), MH (star), MB (diamond) and
CB (crosses).}
\end{figure*}

\end{document}